\documentclass[[12pt,thmsa]{article}
\usepackage{amssymb}

\makeatletter \@addtoreset{equation}{section} \makeatother

\def\ftoday{{\sl {Le \number\day \space\ifcase\month
\or janvier\or f\'evrier\or mars\or avril\or mai \or juin\or
juillet\or ao\^ut\or septembre\or octobre \or novembre \or
d\'ecembre\fi\space \number\year}}}
\def\ptoday{{\sl {\number\day \space de\space \ifcase\month
\or janeiro\or fevereiro\or mar{\c c}o\or abril\or maio \or
junho\or julho\or agosto\or setembro\or outubro \or novembro \or
dezembro\fi\space de\space \number\year}}}
\def\gtoday{{\sl {Den \number\day. \ifcase\month
\or Januar\or Februar\or M\"arz\or April\or Mai \or Juni\or
Juli\or August\or September\or Oktober \or November \or
Dezember\fi\space \number\year}}}
\def\today{{\sl {\ifcase\month
\or January\or February\or March\or April\or May \or June\or
July\or August\or September\or October \or November \or
December\fi \space\number\day,\space
                                            \number\year}}}

\newcommand{\XI}{\XI}

\newcommand{\sla}{\raise.15ex\hbox{$/$}\kern -.57em}
\newcommand{\Sla}{\raise.15ex\hbox{$/$}\kern -.70em}

\newcommand{\complex}{{\kern .1em {\raise .47ex
\hbox {$\scriptscriptstyle |$}}
    \kern -.4em {\rm C}}}
\newcommand{\real}{{{\rm I} \kern -.19em {\rm R}}}
\newcommand{\rational}{{\kern .1em {\raise .47ex
\hbox{$\scripscriptstyle |$}}
    \kern -.35em {\rm Q}}}
\renewcommand{\natural}{{\vrule height 1.6ex width
.05em depth 0ex \kern -.35em {\rm N}}}

\newcommand{\twiddle}{\lower.9ex\rlap{$\kern -.1em\scriptstyle\sim$}}

\newcommand{\eq}{\begin{equation}}
\newcommand{\eqn}[1]{\label{#1}\end{equation}}
\newcommand{\eea}{\end{eqnarray}}
\newcommand{\eqa}{\begin{eqnarray}}
\newcommand{\eqan}[1]{\label{#1}\end{eqnarray}}
\newcommand{\ba}{\begin{array}}
\newcommand{\ea}{\end{array}}
\newcommand{\eqac}{\begin{equation}\begin{array}{rcl}}
\newcommand{\eqacn}[1]{\end{array}\label{#1}\end{equation}}

\begin{document}

{\hfill\large {\bf Matter.tex}\ \ \ptoday} {\hfill\parbox{45mm}{{
hep-th/0312004\\
CBPF-NF-040/03 }} \vspace{3mm}

\begin{center}

{\LARGE\bf Anti-symmetric rank-two Tensor Matter Field on
Superspace for $N_{T}=2$}
\end{center}
\vspace{3mm}

\begin{center}{\large
Wesley Spalenza$^{a,b,}$\footnote{Supported in part by the
Conselho Nacional de Desenvolvimento Cient\'{\i}fico e
Tecnol\'{o}gico CNPq -- Brazil.}, Wander G. Ney$^{a,b,c}$ and J.A.
Helayel-Neto$^{a,b,1}$ } \vspace{1mm}

\noindent $^{a}$Centro Brasileiro de Pesquisas F\'{i}sicas - (CBPF) - Rio de Janeiro - RJ\\
$^{b}$Grupo de F\'{i}sica Te\'{o}rica Jos\'{e} Leite Lopes - (GFT) - Petr\'{o}poles - RJ \\
$^{c}$Centro Federal de Educa\c{c}\~{a}o Tecnol\'{o}gica - (CEFET)
- Campos dos Goytacazes - RJ

\vspace{1mm} E-mails\\
 wesley@cbpf.br, wander@cbpf.br, helayel@cbpf.br.
\vspace{1mm}

\end{center}

\begin{abstract}
In this work, we discuss the interaction between anti-symmetric
rank-two tensor matter and topological Yang-Mills fields. The
matter field considered here is the rank-2 Avdeev-Chizhov tensor
matter field in a suitably extended $N_{T}=2$ SUSY. We start off
from the $N_{T}=2$, $D=4$ superspace formulation and we go over to
Riemannian manifolds. The matter field is coupled to the
topological Yang-Mills field. We show that both actions are
obtained as $Q-$exact forms, which allows us to write the
energy-momentum tensor as $Q-$exact observables.
\end{abstract}

\section{Introduction}

Topological field theories such as Chern-Simons and BF-type gauge
theories probe space-time in its global structure, and this aspect
has a significative relevance in quantum field theories. On the
other hand, there is great deal of interest in anti-symmetric
rank-2 tensor fields that can be put into two categories: gauge
fields or matter fields. In recent years, Avdeev Chizhov
\cite{6,7,8} proposed a model where the antisymmetric tensor
behaves as a matter field.

In a recent work \cite{36}, Geyer-M\"{u}lsch presented a
formulation until then unknown in the literature, which is a
construction of the Avdeev-Chizhov action described in the
topological formalism \cite{27}. This was built for $N_{T}=1$ and
generalized for $N_{T}=2$. Known the properties of the
anti-symmetric rank-two tensor matter field theory, also called
Avdeev-Chizhov field \cite{12}, the supersymmetric properties and
characteristics are presented also in ref. \cite{14}; following
this formalism, we shall write this action in the superfield
formalism, as presented by Horne \cite{33} in topological theories
as a Donaldson-Witten topological theories \cite{19,27}.

Our goal in this work is to discuss the interaction between matter
and topological Yang-Mills fields as presented by Geyer-M\"{u}lsch
\cite{36} for $N_{T}=1$ and $N_{T}=2$. The matter field considered
here is the rank-2 tensor matter field as a complex self-duality
condition \cite{12}. Thus, we write this field now as an
anti-symmetric rank-two tensor matter superfield in $N_{T}=2$ SUSY
in the superspace formalism, founded also in \cite{14}. The matter
field is coupled to the topological Yang-Mills connection by means
of the Blau-Thompson action. We write the Yang-Mills
superconnection as a $2-$superform in a superspace with four
bosonic dimensions spacetime described by Grassmann-odd
coordinates and two fermionic dimensions described by
Grassmann-even coordinates, and them construct the action in a
superfield formalism following the definitions by Horne \cite{33}.
Then, we go over to Riemannian manifolds duely described in terms
of the vierbein and the spin connection, where we take the
gravitation as a background. We introduce and discuss the
Wess-Zumino gauge condition induced by the shift supersymmetry
better detailed in \cite{35}. Then, we arrive at a topological
invariant action as the sum of the Avdeev-Chizhov`s action coupled
to the topological super-Yang-Mills action; both actions are
obtained as $Q-$exact forms, and the energy-momentum tensor is
shown to be $Q-$exact.

\section{The $N_{T}=2$ Super-conection, Super-curvature and Shift Algebra}

Let us now consider the Donaldson-Witten theory, whose space of
solutions is the space of self-dual instantons, $F=*F$. To follow
our superfield formulation, we shall proceed with the definition
of the action of Horne \cite{33} and Blau-Thompson \cite{40,42}.
The $N_{T}=2$ superfield conventions are the ones of \cite{35}.
The superfields superconnection and its associated superghosts are
given as below:
\begin{equation}
\hat{A}=\hat{A}^{a}T_{a},\,\,\hat{C}=\hat{C}^{a}T_{a},
\label{equa1}
\end{equation}
whose the generators belonging the Lie algebra:
\begin{equation}
\lbrack T_{a},T_{b}]=if_{ab}{}^{c}T_{c}.  \label{algLie}
\end{equation}
Expanding the superforms (\ref{equa1}) in component superfields,
we have
\begin{equation}
\hat{A}=A(x_{\mu },\theta ^{I})+E_{I}(x_{\mu },\theta ^{I})d\theta ^{I},\,\,%
\hat{C}=C(x_{\mu },\theta ^{I}),  \label{1.2}
\end{equation}
with $I=1,2$; in component fields, it comes out as below:
\begin{eqnarray}
A(x,\theta ) &=&a(x)+\theta ^{I}\psi _{I}(x)+\frac{1}{2}\theta
^{2}\alpha
(x), \\
E_{I}(x,\theta ) &=&\chi _{I}(x)+\theta ^{I}\phi
_{IJ}(x)+\frac{1}{2}\theta
^{2}\eta _{I}(x), \\
C(x,\theta ) &=&c(x)+\theta ^{I}c_{I}(x)+\frac{1}{2}\theta
^{2}c_{F}(x).
\end{eqnarray}
The associated supercurvature is defined as
\begin{equation}
\hat{F}=\hat{d}\hat{A}+\hat{A}^{2}=(dA+A^{2})+(\partial
_{I}A+D_{A}E_{I})\ d\theta ^{I}+\frac{1}{2}(\partial
_{I}E_{J}+\partial _{J}E_{I}+[E_{I},E_{J}])d\theta ^{I}d\theta
^{J},
\end{equation}
which can also be expressed as: $\hat{F}=F+\Psi _{I}\ d\theta
^{I}+\Phi _{IJ}\ d\theta ^{I}d\theta ^{J}$, whose components read
as follows:
\begin{eqnarray}
F &=&f-\theta ^{I}D_{a}\psi _{I}+\frac{1}{2}\theta ^{2}(D_{a}\alpha +\frac{1%
}{2}\varepsilon ^{IJ}\left[ \psi _{I},\psi _{J}\right] ), \\
\Psi _{I} &=&\psi _{I}+D_{a}\chi _{I}+\theta ^{J}(\varepsilon
_{IJ}\alpha
-\theta ^{J}D_{a}\phi _{IJ}+\theta ^{J}[\psi _{J},\chi _{I}])  \nonumber \\
&&+\theta ^{2}(\frac{1}{2}D_{a}\eta _{I}-\frac{1}{2}\varepsilon
^{KJ}[\psi
_{K},\phi _{IJ}]+\frac{1}{2}[\alpha ,\chi _{I}]), \\
\Phi _{IJ} &=&\frac{1}{2}\{\phi _{IJ}+\phi _{JI}+[\chi _{I},\chi
_{J}]+\theta ^{K}(\varepsilon _{KI}\eta _{J}+\varepsilon _{JK}\eta
_{I}+[\chi _{I},\phi _{JK}]+[\phi _{IK},\chi _{J}])  \nonumber \\
&&+\frac{1}{2}\theta ^{2}([\chi _{I},\eta _{J}]+[\eta _{I},\chi
_{J}]-\varepsilon ^{KL}[\phi _{IK},\phi _{JL}])\},
\end{eqnarray}
where $f=da+a^{2}$ and the covariant derivatives in $a$ being given by $%
D_{a}(\cdot )=d(\cdot )+[a,(\cdot )]$; the symbol $(\cdot )$
represents any field which the derivative act upon. This formalism
with $N_{T}=2,$ it can be found as an example in the work
\cite{COW}.

The SUSY number, $s$, is defined by attributing $-1$ to $\theta $.
Thus, the supersymmetry generators, $Q$, have $s=1$. The BRST
tranformation of the
superconnection (\ref{1.2}) is $s\hat{A}=-\hat{d}\hat{C}-[\hat{A},\hat{C}]=-%
\hat{D}_{\hat{A}}\hat{C}$ and component superfields, is given by
\begin{equation}
\begin{array}{l}
sA=-dC-[A,C]=-D_{A}C, \\
sE_{I}=-\partial _{I}C-[E_{I},C]=-D_{I}C, \\
\,sC=-C^{2},
\end{array}
\label{BRST-transf}
\end{equation}
which in components take the form:
\begin{equation}
\begin{array}{l}
sa=-dc-[a,c]=-D_{a}c\ , \\
s\psi _{I}=-\left[ c,\psi _{I}\right] -D_{a}c_{I}, \\
s\alpha =-[c,\alpha ]-D_{a}c_{F}+\varepsilon ^{IJ}\left[
c_{I},\psi
_{J}\right] , \\
s\chi _{I}=-[c,\chi _{I}]-c_{I}, \\
s\phi _{IJ}=-[c,\phi _{IJ}]-\varepsilon _{IJ}c_{F}+[\chi _{I},c_{J}]\ , \\
s\eta _{I}=-[c,\eta _{I}]-[c_{F},\chi _{I}]+\varepsilon
^{JK}[c_{J},\phi
_{IK}], \\
sc=-c^{2}, \\
sc_{I}=-[c,c_{I}], \\
sc_{F}=-[c,c_{F}]+\frac{1}{2}\varepsilon ^{IJ}\left[
c_{I},c_{J}\right] .
\end{array}
\label{1.7}
\end{equation}
and the super-covariant derivative is decomposed as: $\hat{D}_{\hat{A}%
}=D_{A}+d\theta ^{I}D_{I}$.

The supersymmetry transformations or shift symmetry
transformations are defined as:
\[
Q_{I}A=\partial _{I}A,\,\,Q_{I}E_{J}=\partial
_{I}E_{J},\,\,\,Q_{I}C=\partial _{I}C;
\]
in components, they read as follows:
\begin{equation}
\begin{array}{lll}
Q_{I}a=\psi _{I}\ , & Q_{I}\psi _{J}=-\varepsilon _{IJ}\alpha , &
Q_{I}\alpha =0, \\
Q_{I}\chi _{J}=\phi _{JI}, & Q_{I}\phi _{Jk}=-\varepsilon
_{IK}\eta _{J}\ ,
& Q_{I}\eta _{J}=0, \\
Q_{I}c=c_{I}, & Q_{I}c_{I}=-\varepsilon _{IJ}c_{F}, &
Q_{I}c_{F}=0.
\end{array}
\label{1.8}
\end{equation}

Next, we believe it is interesting to introduce and discuss a sort
of Wess-Zumino gauge choice associated to the shift symmetry
above, which is
the topological BRST transformation. The Wess-Zumino \footnote{%
This name is given since we are dealing with a linear gauge and
scalar ghost field.} gauge seen in \cite{1,35}, is here defined by
the condition
\begin{equation}
\chi _{I}=0\,\,\,\,\,\,and\,\,\,\,\,\phi _{\left[ IJ\right] }=0,
\label{1.9}
\end{equation}
due to the linear shift in the transformations (\ref{1.7}) for
scalar fields $\chi _{I}$ and $\phi _{IJ}$ respectively, with
parameters given by the ghost fields, $c_{I}$ and $c_{F}$.\ There
exists now, only the symmetric field $\phi _{(IJ)}$, that we write
from now on simply as $\phi _{IJ}$. This condition is not
SUSY-invariant under $Q_{I}$, and it can be defined in terms of
the infinitesimal fermionic parameter $\epsilon ^{I}$ as
\[
\widetilde{Q}=\epsilon ^{I}\widetilde{Q}_{I}.
\]
This operator leaves the conditions (\ref{1.9}) invariant, and it
is built up by the combinations of $Q$ with the BRST
transformations in the Wess-Zumino gauge, such that
\begin{equation}
\widetilde{Q}=(s+Q)|_{c_{I}=\varepsilon ^{J}\phi _{IJ},\,\,c_{F}=\frac{1}{2}%
\varepsilon ^{J}\eta _{J}}.  \label{1.10}
\end{equation}
The results in terms of component fields are displayed below:
\begin{equation}
\begin{array}{l}
\widetilde{Q}a=-D_{a}c\ +\epsilon ^{I}\psi _{I}, \\
\widetilde{Q}\psi _{I}=-\left[ c,\psi _{I}\right] -\epsilon
^{J}D_{a}\phi
_{IJ}+\epsilon _{I}\alpha , \\
\widetilde{Q}\alpha =-[c,\alpha ]+\varepsilon ^{IJ}\epsilon
^{K}\left[ \phi
_{Ik},\psi _{J}\right] -\frac{1}{2}\epsilon ^{I}D_{a}\eta _{I}, \\
\widetilde{Q}\phi _{IJ}=-[c,\phi _{IJ}]+\frac{1}{2}\left( \epsilon
_{I}\eta
_{J}+\epsilon _{J}\eta _{I}\right) , \\
\widetilde{Q}\eta _{I}=-[c,\eta _{I}]+\varepsilon ^{JK}\epsilon
^{M}[\phi
_{JM},\phi _{IK}], \\
\widetilde{Q}c=-c^{2}+\epsilon ^{I}\epsilon ^{J}\phi _{IJ}.
\end{array}
\label{1.11}
\end{equation}
in agreement with the transformation found in the works of
\cite{39,42}; the nilpotence reads as
\begin{equation}
(\widetilde{Q})^{2}\varpropto \delta _{\phi _{IJ}},
\label{WZ-gauge}
\end{equation}
that is an infinitesimal transformation of $\phi _{IJ}$. With the
result of the previous section, we are ready to write down the
Blau-Thompson action, which is the invariant Yang-Mills action for
the topological theory.

\section{The Blau-Thompson action}

The associated action for $N_{T}=2$, $D=4$ is the Witten action
\cite {33,39,41}, described in $N_{T}=2$ by the Blau-Thompson
action \cite{40,42}, with gauge completely fixed in terms of the
superfield. For the construction of this action, we wish a
Lagrange multiplier that couples to the topological
super-Yang-Mills so as to manifest its self-duality: $F=*F$. We
then define a $2$-form-superfield Lagrange multiplier, with the
property of anti-self-duality and super-gauge covariant:
$sK=-[C,K]$, such that
\[
K(x,\theta )=k(x)+\theta ^{I}k_{I}(x)+\frac{1}{2}\theta ^{2}\kappa
(x).
\]
We still wish a quadratic term in the last component field of $K$.
Still, we need a $0$-form-superfield to complete the gauge-fixing
for $\Psi _{I}$, which is defined as:
\begin{equation}
H_{I}(x,\theta )=h_{I}(x)+\theta ^{J}h_{JI}(x)+\frac{1}{2}\theta
^{2}\rho _{I}(x).
\end{equation}
To fix the super-Yang-Mills gauge, we define an anti-ghost superfield for $C$%
, being a $0$-form-superfield of fermionic nature
\begin{equation}
\overline{C}(x,\theta )=\overline{c}(x)+\theta ^{I}\overline{c}_{I}(x)+\frac{%
1}{2}\theta ^{2}\overline{c}_{F}(x),  \label{anti-ghost sup}
\end{equation}
we define a $0$-form-superfield Lagrange mulptiplier
\begin{equation}
B(x,\theta )=b(x)+\theta ^{I}b_{I}(x)+\frac{1}{2}\theta ^{2}\beta
(x).
\end{equation}
Their BRST tranformations are $s\overline{C}=B,\,\,sB=0,$ and in
components they reads
\begin{equation}
\begin{array}{lll}
s\overline{c}=b, & s\overline{c}_{I}=b_{I}\,, & s\overline{c}_{F}=\beta , \\
sb=0, & sb_{I}=0, & s\beta =0.
\end{array}
\end{equation}

Therefore the complete Blau-Thompson action in superspace takes
the form

\begin{equation}
S_{BT}=\int d^{2}\theta \,\sqrt{g}\,Tr\{K*F+\zeta K*D_{\theta
}^{2}K+\varepsilon ^{IJ}H_{I}D_{A}*\Psi _{J}+s(\overline{C}d*A)\},
\label{Witten}
\end{equation}
with $\zeta $ being constant. In components, we have
\begin{eqnarray}
S_{BT} &=&\int \sqrt{g}\,Tr\{\frac{1}{2}\kappa *f+\zeta \kappa
*\kappa +\zeta \varepsilon ^{IJ}(k*[\eta _{I},k_{J}]+[k_{J},\eta
_{I}]*k)-\zeta \phi
^{IJ}\phi _{IJ}k*k  \nonumber \\
&&-\frac{1}{2}\varepsilon ^{IJ}k_{I}*D_{a}\psi
_{J}+\frac{1}{2}k*D_{a}\alpha
+\frac{1}{4}k*\varepsilon ^{IJ}[\psi _{I},\psi _{J}]  \nonumber \\
&&+\varepsilon ^{IJ}(\frac{1}{2}\rho _{I}D_{a}*\psi _{J}+\frac{1}{2}%
h_{JI}D_{a}*\alpha -\frac{1}{2}\varepsilon
^{KL}h_{KI}D_{a}*D_{a}\phi _{JL}
\nonumber \\
&&+\frac{1}{2}h_{I}D_{a}*D_{a}\eta _{J}-\varepsilon
^{KL}h_{I}D_{a}*[\psi
_{K},\phi _{JL}]-\frac{1}{2}[h_{I},\psi _{J}]*\alpha  \nonumber \\
&&-\frac{1}{2}\varepsilon ^{KL}[\psi _{K},h_{I}]*D_{a}\phi _{JL}+\frac{1}{2}%
\varepsilon ^{KL}[\psi _{K},h_{LI}]*\psi _{J}+[\alpha ,h_{I}]*\psi
_{J})
\nonumber \\
&&+\frac{1}{2}bd*B+\frac{1}{2}\varepsilon ^{IJ}b_{I}d*\psi _{J}+\frac{1}{2}%
\beta d*a-\frac{1}{2}\overline{c}d*D_{a}c_{F}  \nonumber \\
&&-\frac{1}{2}\varepsilon ^{IJ}\overline{c}d*[\psi _{J},c_{J}]-\frac{1}{2}%
\overline{c}d*[B,c]+\frac{1}{2}\varepsilon
^{IJ}\overline{c}_{I}d*D_{a}c_{J}
\nonumber \\
&&+\frac{1}{2}\varepsilon ^{IJ}\overline{c}_{I}d*[\psi _{J},c]-\frac{1}{2}%
\overline{c}_{F}d*D_{a}c\}.
\end{eqnarray}
where $g$ is the beckground metric of the Riemannian manifold.

In the next section, we shall discuss the Avdeev-Chizhov action in
a general Riemannian manifold with the same background metric.

\section{Tensorial Matter in a General Riemannian Manifold}

To couple the theory above to the Avdeev-Chizhov model, we start
describing the Avdeev-Chizhov action through the complex self-dual
field $\varphi $ \cite{12}, initially written in the 4-dimensional
Minkowskian manifold, whose indices are: $m,n,...$ . We write this
action, according to the work of \cite{12}, as
\begin{equation}
S_{matter}=\int d^{4}x\{(D^{m}\varphi _{mn})^{\dagger
}(D_{p}\varphi ^{pn})+q(\varphi _{mn}^{\dagger }\varphi
^{pn}\varphi ^{\dagger mq}\varphi _{pq})\}.  \label{actionAC}
\end{equation}
Here $q$ is a coupling constant for the self-interaction, and the
covariant derivative $D_{a}^{m}\varphi _{mn}=\partial ^{m}\varphi
_{mn}-[a^{m},\varphi
_{mn}]$; $a^{m}$ is the Lie-algebra-valued gauge potential and we assume $%
\varphi _{mn}$ to belong a given representating of the gauge group
$G$. This action is invariant under the folowing transformations:
\begin{equation}
\delta _{G}(\omega )a_{m}=D_{m}\omega ,\,\,\delta _{G}(\omega
)\varphi _{mn}=\varphi _{mn}\omega ,\,\delta _{G}(\omega )\varphi
_{mn}^{\dagger }=-\omega \varphi _{mn}^{\dagger },
\end{equation}
with $\varphi $ given by
\begin{equation}
\varphi _{mn}=T_{mn}+i\widetilde{T}_{mn},
\end{equation}
which exhibit the properties $\varphi _{mn}=i\widetilde{\varphi }_{mn}$, $%
\widetilde{\widetilde{\varphi }}_{mn}=-\varphi _{mn}$, where the
duality is defined by $\widetilde{\varphi
}_{mn}=\frac{1}{2}\varepsilon _{mnpq}\varphi ^{pq}$.

To treat this theory, in a general Riemannian manifold as a
topological theory, Geyer-M\"{u}lsch \cite{36} rewrite the field
in a four-dimensional Riemannian manifold, endowed of the vierbein
$e_{\mu }^{\,\,\,m}$ and a
spin-connection $\omega _{\mu }^{mn},$ i.e., the tensorial matter read as $%
\varphi _{\mu \nu }=e_{\mu }^{\,\,\,\,m}e_{\nu }^{\,\,\,n}\varphi
_{mn},$ where the action (\ref{actionAC}) is given by
\begin{equation}
S_{matter}=\int d^{4}x\sqrt{g}\{(\nabla _{\mu }\varphi ^{\mu \nu
})^{\dagger }(\nabla _{\rho }\varphi _{\,\,\,\nu }^{\rho
})+q(\varphi _{\mu \nu }^{\dagger }\varphi ^{\rho \nu }\varphi
^{\dagger \mu \lambda }\varphi _{\rho \lambda })\}.  \label{ação
Riemann}
\end{equation}
In this 4-dimensional Riemannian manifold, we find the folowing
properties:
\begin{equation}
\sqrt{g}\varepsilon _{\mu \nu \rho \lambda }\varepsilon
^{mnpq}=e_{[\mu }^{\,\,\,m}e_{\nu }^{\,\,\,n}e_{\rho
}^{\,\,\,p}e_{\lambda ]}^{\,\,\,q},
\end{equation}
\begin{equation}
e_{\mu }^{\,\,\,m}e_{\nu }^{\,\,\,n}g^{\mu \nu }=\eta
^{mn},\,\,e_{\mu }^{\,\,\,m}e_{\nu }^{\,\,\,n}\eta _{mn}=g_{\mu
\nu }.
\end{equation}
The covariant derivative in the Riemannian manifold is now written
in terms of the spin-connection:
\begin{equation}
\nabla _{\mu }=D_{\,\mu }+\omega _{\mu }^{\,\,\,},
\label{cova-derivtive}
\end{equation}
where $\omega _{\mu }^{\,\,\,}=\frac{1}{2}\omega _{\mu
}^{\,\,\,mn}\sigma
_{mn},$ being $\sigma _{mn}$ the generator of the holonomy Euclidean group $%
SO(4)$, also we have: $D_{\mu }=(D_{a})_{\mu },$ where, $a$, is
the Yang-Mills connection.

\section{Supersymmetrization of the Avdeev-Chizhov Action}

From now on, we can write the action (\ref{ação Riemann}) in terms
of superfields, mentioning the conventions of the works
\cite{35,33}. The superfield that accommodates the rank-two
anti-symmetric tensorial matter field, is similar to the one
defined in \cite{14}, being now expressed as a linear fermionic.
This is defined as a rank-two anti-symmetric tensor in the
4-dimensional Riemannian manifold, and with the topological fermionic index $%
I$ referring to the topological SUSY index:
\begin{equation}
\Sigma _{\mu \nu }^{I}(x,\theta )=\lambda _{\mu \nu
}^{I}(x)+\theta ^{I}\varphi _{\mu \nu }(x)+\frac{1}{2}\theta
^{2}\zeta _{\mu \nu }^{I}(x), \label{SuperAC}
\end{equation}
where $\varphi _{\mu \nu }(x)$ is the Avdeev-Chizhov field. The
super-manifold is composed by Riemannian manifold and the
$N_{T}=2$ topological manifold.

The superfield is defined under the SUSY transformations
\begin{equation}
Q_{I}\Sigma _{\mu \nu J}=\partial _{I}\Sigma _{\mu \nu J},
\end{equation}
and in components:
\begin{equation}
\begin{array}{l}
Q_{I}\lambda _{\mu \nu J}=\varepsilon _{IJ}\varphi _{\mu \nu } \\
Q_{I}\varphi _{\mu \nu }=-\zeta _{\mu \nu I} \\
Q_{I}\zeta _{\mu \nu J}=0
\end{array}
\end{equation}

Based on the work of ref. \cite{12}, we rewrite the BRST
transformations, referring the non-Abelian Avdeev-Chizhov model,
in terms of the transformations:
\[
\begin{array}{cc}
s\varphi _{mn}^{i}=ic^{a}(T^{a})^{ij}\varphi _{mn}^{j}, & s\varphi
_{mn}^{\dagger i}=-ic^{a}\varphi _{mn}^{\dagger j}(T^{a})^{ji}, \\
s(\nabla _{m}\varphi _{mn})^{i}=ic^{a}(T^{a})^{ij}(\nabla
_{m}\varphi _{mn})^{j}, & s(\nabla _{m}\varphi _{mn})^{\dagger
i}=-ic^{a}(\nabla _{m}\varphi _{mn})^{\dagger j}(T^{a})^{ji},
\end{array}
\]
where (\ref{algLie}) is the Lie algebra. We wish to write the BRST$-$%
transformation for a supergauge transformation, generalizing the
transformations for the Avdeev-Chizhov fields, according to
\begin{equation}
\begin{array}{l}
s(\Sigma _{\mu \nu }^{I})=iC(\Sigma _{\mu \nu }^{I}), \\
s(\Sigma _{\mu \nu }^{I})^{\dagger }=iC(\Sigma _{\mu \nu
}^{I})^{\dagger };
\end{array}
\end{equation}
in components, we get:
\begin{equation}
\begin{array}{l}
s\lambda _{\mu \nu }^{I}=ic\lambda _{\mu \nu }^{I}, \\
s\lambda _{\mu \nu }^{\dagger I}=-ic\lambda _{\mu \nu }^{\dagger I}, \\
s\varphi _{\mu \nu }=ic\varphi _{\mu \nu }+ic^{I}\lambda _{\mu \nu I}, \\
s\varphi _{\mu \nu }^{\dagger }=-ic\varphi _{\mu \nu }^{\dagger
}-ic^{I}\lambda _{\mu \nu I}^{\dagger }, \\
s\zeta _{\mu \nu }^{I}=ic\zeta _{\mu \nu }^{I}-ic^{I}\varphi _{\mu
\nu
}+ic_{F}\lambda _{\mu \nu }^{I}, \\
s\zeta _{\mu \nu }^{\dagger I}=-ic\zeta _{\mu \nu }^{\dagger
I}+ic^{I}\varphi _{\mu \nu }^{\dagger }-ic_{F}\lambda _{\mu \nu
}^{\dagger I}.
\end{array}
\end{equation}

The super-derivative of the (\ref{SuperAC}) is covariant under the BRST$-$%
transformation, where now, the covariant super-derivative is
\[
{\cal D}_{\mu }(\cdot )=(D_{A})_{\mu }(\cdot )+\omega _{\mu
}(\cdot )=\nabla _{\mu }(\cdot )+\theta ^{I}[\psi _{I\,\mu
},(\cdot )]+\frac{1}{2}\theta ^{2}[\alpha _{\mu },(\cdot )],
\]
acoording to (\ref{cova-derivtive}), then gives
\begin{eqnarray*}
s({\cal D}_{\mu }\Sigma _{\mu \nu }^{I}) &=&C\,({\cal D}_{\mu
}\Sigma _{\mu
\nu }^{I}), \\
s(D_{I}\Sigma _{\mu \nu }^{I}) &=&C\,(D_{I}\Sigma _{\mu \nu
}^{I}),
\end{eqnarray*}
where we chose here, $s\omega _{\mu }=0$.

By now performing BRST$-$transformations on the components that
survive in the $N_{T}=2$ Wess-Zumino gauge (\ref{1.10}), we find:
\begin{equation}
\begin{array}{l}
\widetilde{Q}\lambda _{\mu \nu I}=\epsilon ^{J}\varepsilon
_{JI}\varphi
_{\mu \nu }+ic\lambda _{\mu \nu I}, \\
\widetilde{Q}\lambda _{\mu \nu I}^{\dagger }=\epsilon
^{J}\varepsilon
_{JI}\varphi _{\mu \nu }^{\dagger }-ic\lambda _{\mu \nu I}^{\dagger }, \\
\widetilde{Q}\varphi _{\mu \nu }=ic\varphi _{\mu \nu }+i\epsilon
^{I}\zeta
_{\mu \nu I}+i\epsilon ^{I}\phi _{IJ}\lambda _{\mu \nu }^{J}, \\
\widetilde{Q}\varphi _{\mu \nu }^{\dagger }=-ic\varphi _{\mu \nu
}^{\dagger }-i\epsilon ^{I}\zeta _{\mu \nu I}^{\dagger }-i\epsilon
^{I}\phi
_{IJ}\lambda _{\mu \nu }^{\dagger J}, \\
\widetilde{Q}\zeta _{\mu \nu I}=ic\zeta _{\mu \nu I}-i\epsilon
^{J}\phi
_{JI}\varphi _{\mu \nu }+i\epsilon ^{J}\eta _{J}\lambda _{\mu \nu I}, \\
\widetilde{Q}\zeta _{\mu \nu I}^{\dagger }=-ic\zeta _{\mu \nu
I}^{\dagger }+i\epsilon ^{J}\phi _{JI}\varphi _{\mu \nu }^{\dagger
}-i\epsilon ^{J}\eta _{J}\lambda _{\mu \nu I}^{\dagger },
\end{array}
\label{transfgaugeTopo}
\end{equation}
in agreement to (\ref{WZ-gauge}).

We build up rank-two anti-symmetric tensorial matter field in a
superspace formulation, leaving the superfield with the same
properties as shown in \cite{14}; this is invariant under gauge
transformations (\ref {transfgaugeTopo}) and SUSY transformations.
The kinetic term is proposed as
\[
S_{kin}=\int d^{4}xd^{2}\theta \sqrt{g}\varepsilon ^{IJ}\{({\cal
D}_{\mu }\Sigma _{I}^{\mu \nu })^{\dagger }({\cal D}_{\rho }\Sigma
_{\,\,\,\nu \,J}^{\rho })\}.
\]
In components, we get:
\begin{eqnarray}
S_{kin} &=&\int d^{4}x\sqrt{g}\{\frac{1}{2}(\nabla _{\mu }\varphi
^{\mu \nu
})^{\dagger }(\nabla _{\rho }\varphi _{\,\,\,\nu \,}^{\rho })+\frac{1}{2}%
\varepsilon ^{IJ}(\nabla _{\mu }\lambda _{I}^{\mu \nu })^{\dagger
}(\nabla
_{\rho }\zeta _{\,\,\,\nu \,J}^{\rho })  \nonumber \\
&&+\frac{1}{2}\varepsilon ^{IJ}(\nabla _{\mu }\zeta _{I}^{\mu \nu
})^{\dagger }(\nabla _{\rho }\lambda _{\,\,\,\nu \,J}^{\rho
})+(\nabla _{\mu }\varphi ^{\mu \nu })^{\dagger }[\psi _{\rho
}^{I},\lambda _{\,\,\,\nu
\,I}^{\rho }]  \nonumber \\
&&+[\psi _{\mu \,J},\varphi ^{\dagger \mu \nu }](\nabla _{\rho
}\varphi _{\,\,\,\nu \,}^{\rho })+\varepsilon ^{IJ}(\nabla _{\mu
}\lambda _{I}^{\mu \nu })^{\dagger }\left( [\alpha _{\rho
},\lambda _{\,\,\,\nu \,J}^{\rho
}]+[\psi _{\rho \,J},\varphi _{\,\,\,\nu }^{\rho }]\right)  \nonumber \\
&&+\varepsilon ^{IJ}\left( [\alpha _{\mu },\lambda _{I}^{\dagger
\mu \nu }]+[\psi _{\mu \,J},\varphi ^{\dagger \mu \nu }]\right)
(\nabla _{\rho }\lambda _{\,\,\,\nu \,J}^{\rho })\}
\end{eqnarray}

The interaction term has the peculiarity of presenting two
derivatives of the Grassmann coordinates; it should also be
invariant under the gauge transformations (\ref{transfgaugeTopo})
and supersymmetry. We write it as
\begin{equation}
S_{int}=\int d^{4}xd^{2}\theta \sqrt{g}\{\varepsilon
^{IJ}\varepsilon ^{LM}(\Sigma _{\mu \nu \,I})^{\dagger
}D^{K}(\Sigma _{J}^{\rho \nu })(\Sigma _{L}^{\mu \lambda
})^{\dagger }D_{K}(\Sigma _{\rho \lambda \,M})\}
\end{equation}
where $D_{K}(\cdot )=\partial _{K}(\cdot )+[E_{K},(\cdot )];$ in
components,
\begin{eqnarray}
S_{int} &=&\frac{1}{2}\int d^{4}x\sqrt{g}\{\varphi _{\mu \nu
}^{\dagger }\varphi ^{\rho \nu }\varphi ^{\dagger \mu \lambda
}\varphi _{\rho \lambda }-\varepsilon ^{IJ}[(\lambda _{\mu \nu
\,I}^{\dagger }\zeta _{J}^{\rho \nu }+\zeta _{\mu \nu
\,I}^{\dagger }\lambda _{J}^{\rho \nu })\varphi ^{\dagger
\mu \lambda }\varphi _{\rho \lambda }  \nonumber \\
&&-\varphi _{\mu \nu }^{\dagger }\varphi ^{\rho \nu }(\lambda
_{I}^{\dagger \mu \lambda }\zeta _{\rho \lambda \,J}+\zeta
_{I}^{\dagger \mu \lambda }\lambda _{\rho \lambda
\,J})]+\varepsilon ^{IJ}\varepsilon ^{KL}[\lambda _{\mu \nu
\,I}^{\dagger }\zeta _{J}^{\rho \nu }(\lambda _{K}^{\dagger \mu
\lambda }\zeta _{\rho \lambda \,L}+\zeta _{K}^{\dagger \mu \lambda
}\lambda
_{\rho \lambda \,L})  \nonumber \\
&&+\zeta _{\mu \nu \,I}^{\dagger }\lambda _{J}^{\rho \nu }(\lambda
_{K}^{\dagger \mu \lambda }\zeta _{\rho \nu \,L}+\zeta
_{K}^{\dagger \mu \lambda }\lambda _{\rho \nu \,L})+\lambda _{\mu
\nu \,I}^{\dagger }\lambda _{J}^{\rho \nu }[\eta _{L},\lambda
_{K}^{\mu \lambda }]\varphi _{\rho
\lambda } \\
&&-\lambda _{\mu \nu \,I}^{\dagger }\varphi _{J}^{\rho \nu }\eta
_{L}\lambda _{K}^{\dagger \mu \lambda }\lambda _{\rho \lambda
}+\varphi _{\mu \nu \,}^{\dagger }\lambda _{J}^{\rho \nu }\eta
_{I}\lambda _{K}^{\dagger \mu \lambda }\lambda _{\rho \lambda
\,L}-\lambda _{\mu \nu \,I}^{\dagger }\lambda _{J}^{\rho \nu }\eta
_{L}\lambda _{K}^{\dagger \mu \lambda }\varphi
_{\rho \lambda } \\
&&-\lambda _{\mu \nu \,I}^{\dagger }\lambda _{J}^{\rho \nu }\eta
_{K}\varphi ^{\dagger \mu \lambda }\lambda _{\rho \lambda
\,L}+\lambda _{\mu \nu \,I}^{\dagger }\lambda _{J}^{\rho \nu }\phi
^{MN}\phi _{MN}\lambda _{K}^{\dagger \mu \lambda }\lambda _{\rho
\lambda \,L}]\}.
\end{eqnarray}

The total action is being determinad for: $S_{Kin}+qS_{Int}$, such
that
\begin{equation}
S_{AC}=-\int d^{2}\theta \sqrt{g}\{\varepsilon ^{IJ}({\cal D}_{\mu
}\Sigma _{I}^{\mu \nu })^{\dagger }({\cal D}_{\rho }\Sigma
_{\,\,\,\nu \,J}^{\rho })+q\varepsilon ^{IJ}\varepsilon
^{LM}(\Sigma _{\mu \nu \,I})^{\dagger }D^{K}(\Sigma _{J}^{\rho \nu
})(\Sigma _{L}^{\mu \lambda })^{\dagger }D_{K}(\Sigma _{\rho
\lambda \,M})\},
\end{equation}
where $q$ is a quartic coupling constant. In components, we have
the Avdeev-Chizhov action plus its partness:
\begin{eqnarray}
S_{AC} &=&\int d^{4}x\sqrt{g}\{\frac{1}{2}(\nabla _{\mu }\varphi
^{\mu \nu
})^{\dagger }(\nabla _{\rho }\varphi _{\,\,\,\nu \,}^{\rho })+\frac{1}{2}%
\varepsilon ^{IJ}(\nabla _{\mu }\lambda _{I}^{\mu \nu })^{\dagger
}(\nabla
_{\rho }\zeta _{\,\,\,\nu \,J}^{\rho })  \nonumber \\
&&+\frac{1}{2}\varepsilon ^{IJ}(\nabla _{\mu }\zeta _{I}^{\mu \nu
})^{\dagger }(\nabla _{\rho }\lambda _{\,\,\,\nu \,J}^{\rho
})+(\nabla _{\mu }\varphi ^{\mu \nu })^{\dagger }[\psi _{\rho
}^{I},\lambda _{\,\,\,\nu
\,I}^{\rho }]  \nonumber \\
&&+[\psi _{\mu \,J},\varphi ^{\dagger \mu \nu }](\nabla _{\rho
}\varphi _{\,\,\,\nu \,}^{\rho })+\varepsilon ^{IJ}(\nabla _{\mu
}\lambda _{I}^{\mu \nu })^{\dagger }\left( [\alpha _{\rho
},\lambda _{\,\,\,\nu \,J}^{\rho
}]+[\psi _{\rho \,J},\varphi _{\,\,\,\nu }^{\rho }]\right)  \nonumber \\
&&+\varepsilon ^{IJ}\left( [\alpha _{\mu },\lambda _{I}^{\dagger
\mu \nu }]+[\psi _{\mu \,J},\varphi ^{\dagger \mu \nu }]\right)
(\nabla _{\rho
}\lambda _{\,\,\,\nu \,J}^{\rho })  \nonumber \\
&&+q(\varphi _{\mu \nu }^{\dagger }\varphi ^{\rho \nu }\varphi
^{\dagger \mu \lambda }\varphi _{\rho \lambda }-\varepsilon
^{IJ}[(\lambda _{\mu \nu \,I}^{\dagger }\zeta _{J}^{\rho \nu
}+\zeta _{\mu \nu \,I}^{\dagger }\lambda _{J}^{\rho \nu })\varphi
^{\dagger \mu \lambda }\varphi _{\rho \lambda }
\nonumber \\
&&-\varphi _{\mu \nu }^{\dagger }\varphi ^{\rho \nu }(\lambda
_{I}^{\dagger \mu \lambda }\zeta _{\rho \lambda \,J}+\zeta
_{I}^{\dagger \mu \lambda }\lambda _{\rho \lambda
\,J})]+\varepsilon ^{IJ}\varepsilon ^{KL}[\lambda _{\mu \nu
\,I}^{\dagger }\zeta _{J}^{\rho \nu }(\lambda _{K}^{\dagger \mu
\lambda }\zeta _{\rho \lambda \,L}+\zeta _{K}^{\dagger \mu \lambda
}\lambda
_{\rho \lambda \,L})  \nonumber \\
&&+\zeta _{\mu \nu \,I}^{\dagger }\lambda _{J}^{\rho \nu }(\lambda
_{K}^{\dagger \mu \lambda }\zeta _{\rho \nu \,L}+\zeta
_{K}^{\dagger \mu \lambda }\lambda _{\rho \nu \,L})+\lambda _{\mu
\nu \,I}^{\dagger }\lambda _{J}^{\rho \nu }[\eta _{L},\lambda
_{K}^{\mu \lambda }]\varphi _{\rho
\lambda }  \nonumber \\
&&-\lambda _{\mu \nu \,I}^{\dagger }\varphi _{J}^{\rho \nu }\eta
_{L}\lambda _{K}^{\dagger \mu \lambda }\lambda _{\rho \lambda
}+\varphi _{\mu \nu \,}^{\dagger }\lambda _{J}^{\rho \nu }\eta
_{I}\lambda _{K}^{\dagger \mu \lambda }\lambda _{\rho \lambda
\,L}-\lambda _{\mu \nu \,I}^{\dagger }\lambda _{J}^{\rho \nu }\eta
_{L}\lambda _{K}^{\dagger \mu \lambda }\varphi
_{\rho \lambda }  \nonumber \\
&&-\lambda _{\mu \nu \,I}^{\dagger }\lambda _{J}^{\rho \nu }\eta
_{K}\varphi ^{\dagger \mu \lambda }\lambda _{\rho \lambda
\,L}+\lambda _{\mu \nu \,I}^{\dagger }\lambda _{J}^{\rho \nu }\phi
^{MN}\phi _{MN}\lambda _{K}^{\dagger \mu \lambda }\lambda _{\rho
\lambda \,L}])\}.
\end{eqnarray}
It is invariant under conformal transformations. Therefore, the
total gauge invariant action can be written as: $S_{AC}+S_{BT}$.
We could also have replace $S_{BT}$ by the super$-BF$ action
described in the work of ref. \cite {COW}.

The $Q-$exactness of the total action above is also true for
$N_{T}=2$ SUSY
as in \cite{36}; this is so because the fermionic volume element $%
Q^{2}\propto Q_{1}Q_{2}$ , which means the exactness in the charge $Q_{1}$, $%
Q_{2}$ of this action. This proof for $N_{T}=1$ and general
$N_{T}$ , is given in the works \cite{35}, where the total action
is also $s-$exact. According to Blau-Thompson in their review
\cite{34}, the energy-momentum tensor $\Theta _{\mu \nu }$ is also
$Q-$exact,
\begin{equation}
{\cal O}=\langle 0|\Theta _{\mu \nu }|0\rangle =\langle 0|\frac{2}{\sqrt{g}}%
\frac{\delta }{\delta g^{\mu \nu }}(S_{BT}+S_{AC})|0\rangle
=\langle 0|Q\,\Upsilon _{\mu \nu }|0\rangle
\end{equation}
ensuring the topological nature of the theory, where we shall just
use the Avdeev-Chizhov kinetic term, because the interaction term
carries the coupling constant $q$, which is irrelevant for the
attainment of the observables of the theory \cite{36}.

\section*{Concluding Remarks}

The main goal of this paper is the settlement of a topological
superspace formulation for the investigation of the coupling
between the rank-two Avdeev-Chizhov matter field and Yang-Mills
fields. It comes out that the stress tensor is $Q-$exact. This
opens us the way for the identification of a whole class of
obsevables that we are trying to classify \cite{W}.

It is worthwhile to draw the attention here to the shift symmetry
that allows us to detect the ghost caracter of the Avdeev-Chizhov
field. On the other hand, it is known that there appears a ghost
mode in the spectrum of excitations of our tensor matter field
\cite{6}. The connection between these two observations remain to
be clarified. The fact that the Avdeec-Chizhov field manifest
itself as a ghost guide future developments in the quest for a
consistent mechanism to systematically decouple the unphysical
mode mentioned above.

We are also trying to embed the tensor field in the framework of a
gauge theory with Lorentz symmetry breaking \cite{WWH}. We expect
that this breaking may identify the right ghost mode present among
the two spin $1$ components of the Avdeev-Chizhov field.

{\bf Acknowledgments.} We thank \'{A}lvaro Nogueira, Clisthenis P.
Constantinidis, Jos\'{e}.L. Boldo, Daniel H.T. Franco for many
useful discussions and the Prof. Olivier Piguet for the the great
help and
encouragement. 

\appendix

\section*{Appendices}

\section{Conventions}

The topological fermionic index: $I=1,2$, is lowered and raised by
the anti-symmetric Levi-Civita tensor: $\varepsilon _{IJ}$,
$\varepsilon ^{IJ}$, with $\varepsilon ^{12}=-\varepsilon
_{12}=1$. The $\theta -$coordinates definitions: $\theta
^{I}=\varepsilon ^{IJ}\theta _{J},\,\,\theta _{I}=\varepsilon
_{IJ}\theta ^{J},\,\,$the quadratic forms are:
\[
\theta ^{2}=\theta ^{I}\theta _{I}=-\theta _{I}\theta
^{I},\,\,\,\theta ^{I}\theta ^{J}=-\frac{1}{2}\varepsilon
^{IJ}\theta ^{2},\,\,\theta _{I}\theta _{J}=\frac{1}{2}\varepsilon
_{IJ}\theta ^{2},
\]
with $\varepsilon _{IK}\varepsilon ^{KJ}=\delta _{I}^{\,\,J}$. The
derivatives in the $\theta -$coordinates are defined by
\begin{equation}
\partial _{I}=\frac{\partial }{\partial \theta ^{I}},\,\,\,\partial ^{I}=%
\frac{\partial }{\partial \theta _{I}}\,\,\,and\,\,\,\,\partial
_{I}\theta ^{J}\stackrel{Def}{=}\delta _{I}^{\,\,J}.
\end{equation}
thus we have

\[
\partial _{I}f(x,\theta )=\varepsilon _{IJ}\partial ^{J}f(x,\theta ),
\]
with $f(x,\theta )$ a any superfunction. Deriving the $\theta
-$coordinates gives
\begin{equation}
\partial ^{I}\theta ^{J}=-\varepsilon ^{IJ},\,\,\,\,\partial _{I}\theta
_{J}=-\varepsilon _{IJ}
\end{equation}

A superfield is expanded as: $F(x,\theta )=f(x)+\theta ^{I}f_{I}(x)+\frac{1}{%
2}\theta ^{2}f_{F},$ obeying the transformation $Q_{I}F(x,\theta )\stackrel{%
Def}{=}\partial _{I}F(x,\theta ).$ In components, we have:

\begin{equation}
Q_{I}f=f_{I}\,\,;\,\,\,Q_{I}f_{J}=-\varepsilon
_{IJ}f_{F}\,\,;\,\,\,Q_{I}f_{F}=0.
\end{equation}

Characteristics table of the superconnection fields:

\begin{equation}
\begin{tabular}{|c|c|c|c|c|c|c|c|c|c|c|}
\hline
$Charge\backslash \,\,fields$ & $\epsilon ^{I}$ & $a$ & $\psi ^{I}$ & $%
\alpha $ & $\chi ^{I}$ & $\phi ^{IJ}$ & $\eta ^{I}$ & $c$ &
$c^{I}$ & $c_{F}$
\\ \hline
$s$ & $-1$ & $0$ & $1$ & $2$ & $1$ & $2$ & $3$ & $0$ & $1$ & $2$
\\ \hline $g$ & $1$ & $0$ & $0$ & $0$ & $0$ & $0$ & $0$ & $1$ &
$1$ & $1$ \\ \hline $p$ & $0$ & $1$ & $1$ & $1$ & $0$ & $0$ & $0$
& $0$ & $0$ & $0$ \\ \hline
$P_{grs}$ & $+$ & $-$ & $+$ & $-$ & $-$ & $+$ & $-$ & $-$ & $+$ & $-$ \\
\hline
\end{tabular}
\end{equation}
\newline
where $s$: susy number, $g$: ghost number, $p$: degree form,
$P_{grs}$: Grassmann parity.

\section{Rules for Topological Grassmannian integration}

The definition of integration in this topological SUSY
representation is
\begin{equation}
\int d\theta ^{I}\stackrel{Def}{=}\partial _{I}.
\end{equation}
This result is applied to a superfunction $f(x,\theta )$, so that
the volume element is
\begin{equation}
\int d^{2}\theta f(x,\theta
)\stackrel{Def}{=}\frac{1}{4}\varepsilon ^{IJ}\partial
_{I}\partial _{J}f(x,\theta );
\end{equation}
therefore, the square of the supersymmetric charge operator (shift
operator) is defined by:
\[
Q^{2}=Q^{I}Q_{I}=\partial ^{I}\partial _{I}=4\int d^{2}\theta ,
\]
which is a volume element too.


\end{document}